\documentclass[aps,pra,showpacs,floatfix,reprint]{revtex4-1}
\usepackage{amsfonts}
\usepackage{amsmath}
\usepackage{amssymb}
\usepackage{latexsym}
\usepackage{graphicx}
\usepackage{epstopdf}
\usepackage{hyperref}

\begin{document}

\title{Decoherent time-dependent transport beyond the Landauer-B\"{u}ttiker formulation: A
quantum-drift alternative to quantum jumps}
\author{Lucas J. Fern\'andez-Alc\'azar$^{1}$}
\email{lfernan4@famaf.unc.edu.ar}
\author{ Horacio M. Pastawski$^{1}$}
\affiliation{$^{1}$Instituto de F\'{\i}sica Enrique Gaviola and Facultad de Matem\'{a}tica , 
Astronom\'{\i}a y F\'{\i}sica, Universidad Nacional de C\'{o}rdoba,
Ciudad Universitaria, C\'{o}rdoba, 5000, Argentina,}

\begin{abstract}
We develop and implement a model for decoherence in time-dependent
transport. Inspired in a dynamical formulation of the Landauer-B\"{u}ttiker
equations, it boils down into a form of wave function that undergoes a
smooth stochastic drift of the phase in a local basis, the quantum-drift
(QD) model. This drift is nothing else but a local energy fluctuation.
Unlike quantum-jumps (QJ) models, no jumps are present in the density as the
evolution is unitary. As a first application, we address the transport
through a resonant state $\left\vert 0\right\rangle $ that undergoes
decoherence. Its numerical resolution shows the equivalence with the
decoherent steady-state transport in presence of a B\"{u}ttiker's voltage
probe. In order to test the dynamics we consider two many-spin systems,
which are cases of experimental interest, where a local energy fluctuation
is a natural phenomenon. A two-spin system is reduced to a two-level system
(TLS) that oscillates among $\left\vert 0\right\rangle $ $\equiv $ $%
\left\vert \uparrow \downarrow \right\rangle $ and $\left\vert
1\right\rangle \equiv $ $\left\vert \downarrow \uparrow \right\rangle $. We
show that the QD model recovers not only the exponential damping of the
oscillations in the low perturbation regime, but also the non-trivial
bifurcation of the damping rates at a critical point, i.e., the quantum
dynamical phase transition. We also address the spin-wave-like dynamics of
local polarization in a spin chain. By averaging over $N_{\mathbf{s}}$
realizations, the QD solution has about half the dispersion respect to the
mean dynamics than QJ. By evaluating the Loschmidt echo (LE), we find that
the pure states $\left\vert 0\right\rangle $ $\equiv $ $\left\vert \uparrow
\downarrow \right\rangle $ and $\left\vert 1\right\rangle \equiv $ $%
\left\vert \downarrow \uparrow \right\rangle $ are quite robust against the
local decoherence. In contrast, the LE, and hence coherence, decays faster
when the system is in a superposition state $\left( \left\vert \uparrow
\downarrow \right\rangle \pm \left\vert \downarrow \uparrow \right\rangle
\right) /\sqrt{2}$, which is consistent with the general trend recently
observed in spin systems through NMR. Because of its simple implementation, the
method is well suited to assess decoherent transport problems as well as to
include decoherence in both one-body and many-body dynamics.
\end{abstract}
\maketitle

\section{Introduction}

The last decade has seen an increasing demand to describe quantum dynamics
on a variety of complex systems in the presence of an environment. Among them
are atomic systems in optical lattices \cite{Bloch04,CirZ04}, networks of
interacting spins \cite{CRC07} and charge and magnetization dynamics of
nanoscopic devices in a transport set up \cite{LJPWR03Nat,SKVCirGi14,PLMHG05}.

The most common way to deal with environmental decoherence in small closed
systems, is the master equation for the density matrix in a Lindblad form 
\cite{Lin76,Koss72}. In order to deal with bigger systems, the most standard
approaches implement the Redfield theory for a relaxation superoperator
which does not ensure the strict unitarity of the Lindbad form \cite%
{Redfield,Slichter}. Alternatively, some works pointed to strategies for
computing the evolution of an open system based on the stochastic dynamics
of a state vector that suffers instantaneous quantum jumps (QJ) \cite%
{DCM92,MCD93,TM92,Car89,Car93}. Indeed, for large systems, it was shown that
the stochastic method is faster than the density matrix implementations \cite%
{Petr97}.

With regard to decoherent steady state transport, the traditional
evaluation in terms of the Kubo linear response \cite{Alt85,Th74} was
progressively replaced by the Landauer's motto that \textit{conductance is a
quantum transmittance} \cite{IL99}. This is implemented in the Landauer-B%
\"{u}ttiker (LB) scattering formulation where decoherent processes are
induced by current conserving voltmeters. These impose a self-consistent
reinjection of the electrons that ensures the current cancellation. B\"{u}%
ttiker had the idea that this reinjection describes a decoherent process 
\cite{Bt86PRB}, a concept extended by D'Amato and Pastawski to deal with
more general situations \cite{DP90,GLBE1} becoming a popular tool to address
decoherent transport \cite{ZimG08,NozRPC12,MTDP14}. A dynamical formulation,
called the generalized Landauer-B\"{u}ttiker equation (GLBE), based on the Kadanoff-Baym-Keldysh (KBK) quantum field theory
for non-equilibrium processes was then developed \cite{GLBE2}. The GLBE seeks to
find, in a linear response, the non-equilibrium Keldysh density function
which is proportional to the density matrix. In recent years, there has been
a burst of progress in the use of the KBK formulation of non-equilibrium
dynamics in a framework consistent with \textit{ab initio} calculations \cite%
{SvL09,SvL12}. Nonetheless, numerical solutions of many-body systems become
excessively demanding. In particular, they involve time integrals of
self-consistent memory kernels \cite{SvL09,SvL12,GLBE2}. Additionally, in
strongly interacting many-body systems, which are beyond a mean-field
description, such as spin systems, it would involve costly averages over the
participating configurations.

The above limitations can be overcome by resorting to two strategies. The
Trotter-Suzuki step-by-step evolution makes practical some calculations of
the Keldysh density function \cite{ADLP07,DALP07(ssc)}. The ensemble average
is bypassed by a recent strategy, dubbed \textquotedblleft quantum
parallelism\textquotedblright , which uses a \textit{single wave function}
that is a superposition of all the states participating in the statistical
ensemble \cite{ADLP08}. However, these strategies are limited to coherent
dynamics.

In this work, we propose a stochastic model that extends the B\"{u}%
ttiker-D'Amato-Pastawski approach to evaluate decoherent time-dependent
problems. Our quantum-drift (QD) model is based on a wave function
stochastic dynamics. Within a discrete-time set up, we impose an incoherent
reinjection that ensures density conservation. Then, as in quantum
parallelism, we propose that the wave function should sum up both a coherent
and an incoherent part. Thus, being fairly representative of a set of
stochastic interaction \textquotedblleft histories\textquotedblright , this
wave function does not present jumps but smooth drifts in a single unitary
evolution. Further averages over realizations are performed only in the
needed amount according to the addressed observable.

This paper is organized as follows. In Sec. II we review the basis of the B%
\"{u}ttiker's model for transport in open systems. In Sec. III we present
the basis of our QD model. By addressing a wave packet dynamics through a
double barrier resonant tunneling device (DBRTD), in Sec. IV, we show that
QD recovers the decoherent steady state transmittances of the B\"{u}ttiker
formulation. In Sec. V we compare the QD dynamics with the Keldysh solution
in a decoherent closed system with a simple but non-trivial situation: a two-level system (TLS) that undergoes a quantum dynamical phase transition
(QDPT) \cite{ADLP06}. In Sec. VI, we perform a many-spin calculation for a
case of experimental and theoretical interest, that of a quantum channel,
showing the agreement of QD and QJs and comparing their numerical
performance. Finally, in Sec. VII we use the QD model to evaluate
decoherence through the Loschmidt echo (LE) in the TLS of the Sec. V
under local decoherence processes \cite{JalP01}. This allows us to show that
while Rabi oscillations decay uniformly, decoherence is not uniform: it
affects the system only while it goes through non-local superposition. We
present the final discussion in Sec. VII.

\section{Decoherent quantum transport: reinjection, parallelism,
attenuation, and energy broadening.}

The first phenomenological model for decoherence was developed for  B\"{u}%
ttiker in the context of electronic transport in phase-coherent mesoscopic
systems \cite{IL99}. He realized that decoherence may be introduced by
including a terminal connected to a voltmeter. This key idea may be readily
visualized by considering a three-terminal circuit. There, two terminals
(source $L$ and drain $R$) provide an infinitesimal voltage difference that
produce a current through the system while the third one, $\phi $, is
connected to the voltmeter. The electrons from the left ($L$) and right ($R$%
) leads that enter into the voltmeter undergo a decoherent process. An
appropriate local chemical potential at $\phi $ (i.e. the measured voltage)
ensures current cancellation, i.e. a \textit{reinjected} electron
compensates each electron that flies into the voltmeter. Then, this electron
does not keep any memory or phase correlation with respect to the electrons inside
the sample.

Let consider that $T_{ij}$ represents the quantum coherent transmittance
from the $j$ to $i$ channel connecting them to the reservoirs ($i\neq j,$
take the values $L,R$ or $\phi $). The application of the Landauer-B\"{u}%
ttiker equations for a system with one voltmeter results in a transmittance
through the system given by%
\begin{equation}
\widetilde{T}_{RL}=\underset{coherent}{\underbrace{T_{RL}}}+\underset{%
decoherent}{\underbrace{\frac{T_{R\phi }T_{\phi L}}{T_{R\phi }+T_{\phi L}}}}.
\label{eq_decoherent_transmittance}
\end{equation}%
The first term is the probability that the particle travels from $L$ to $R$
without undergoing a decoherent process at the voltmeter $\phi $. The second
term accounts for those electrons that have interacted with the environment
at $\phi $. Indeed, one can recognize it as the conductance of two \textit{%
parallel pathways}. One of them with a conductance $(2e/h)T_{RL}$, while the
other one adds a series of two conductances, $(2e/h)T_{R\phi }$ and $%
(2e/h)T_{\phi L}$.

These results adopt a more concrete form by introducing an explicit model.
Let us consider a quantum dot where $E_{0}$ is the relevant eigenstate
energy, $\varepsilon $ the Fermi energy, $\overline{E_{0}}$ the local energy
properly shifted by the presence of the contacts, and $\Gamma _{L}$ and $%
\Gamma _{R}$, the energy uncertainties produced by the escape toward the
left and right leads, respectively. The dot's Green's function is defined in
terms of the effective Hamiltonian parameters as: $G(\varepsilon )=\left[
(\varepsilon -\overline{E_{0}})+\mathrm{i}(\Gamma _{0}+\Gamma _{\phi })%
\right] ^{-1}$, where $\Gamma _{0}=\Gamma _{L}+\Gamma _{R}$ is the natural
width due to the presence of the leads \cite{PM01,GLBE2}. Thus, each of the
transmittances used above may be written explicitly in terms of $%
G(\varepsilon )$ by using the Fisher-Lee formula $T_{ij}(\varepsilon
)=2\Gamma _{i}\left\vert G(\varepsilon )\right\vert ^{2}2\Gamma _{j}$ with $%
i\neq j$. \ If $T_{RL}^{(0)}$ is the transmittance from $L$ to $R$ that
accounts for electrons in absence of the voltmeter (i.e. when $\Gamma _{\phi
}=0$), the coherent part $T_{RL}$ in Eq. \ref{eq_decoherent_transmittance}
can be further written as the product of $T_{RL}^{(0)}$ and an \textit{%
attenuation} factor $\left( 1-\Lambda (\varepsilon )\right) $ \cite%
{attenuation}. Thus, the effective transmittance is written in terms of a
attenuated coherent part plus an incoherent one.

D'Amato and Pastawski's (DP) model generalize these ideas and introduce an
effective Hamiltonian that constitutes a microscopic model for decoherence
in the steady state \cite{DP90}. In this case, the isolated system is
described by a Hamiltonian $H_{0}$. DP identify the escape of the electrons
towards the voltmeter with their interaction with the infinite degrees of
freedom of an environment. Here, decoherence is induced by local processes
(e.g., a voltage probe, a local phonon bath) in the Fermi golden rule (FGR)
approximation. These interactions produce an energy uncertainty $\Gamma
_{\phi }$ for each local state with a rate of system-environment
interaction, $1/\tau _{SE}=2\Gamma _{\phi }/\hbar $, which has an
irreversible character \cite{DP90,GLBE2,PM01}.

The local density of states (LDoS) is calculated from the dot's Green's
function

\begin{equation}
N_{0}(\varepsilon )=-\frac{1}{\pi }\mathit{Im}G^{(0)}(\varepsilon )=\frac{1}{%
\pi }\frac{\Gamma _{0}}{(\varepsilon -\overline{E}_{0})^{2}+\Gamma _{0}^{2}}.
\end{equation}%
By including the system-environment interaction, the LDoS acquire an extra
energy uncertainty or \textit{broadening} $\Gamma _{\phi }$. Then, the LDoS
in presence of decoherence, $\widetilde{N}_{0}(\varepsilon )$, is obtained
from $N_{0}(\varepsilon )$ by replacing the characteristic width, $\Gamma
_{0}$, by $\Gamma _{0}+\Gamma _{\phi }$. On the other hand, $\widetilde{N}%
_{0}$ may be obtained by considering that individual decoherent processes
shifts the resonances in an amount $\Delta E$ from $\overline{E}_{0}$. Then,
by averaging over the possible $\Delta E$, 
\begin{eqnarray}
\widetilde{N}_{0}(\varepsilon ) &=&\int N_{0}(\varepsilon -\Delta E)P(\Delta
E)\mathrm{d}\Delta E,  \label{Eq_Lorentzian-convolution} \\
\text{with }P(\Delta E) &=&\frac{1}{\pi }\frac{\Gamma _{\phi }}{\left(
\Delta E\right) ^{2}+\Gamma _{\phi }{}^{2}},
\end{eqnarray}%
where, $P(\Delta E)$ is a Lorentzian probability distribution for the shifts 
$\Delta E$. Similar broadening occurs in other observables that depend on
energy, such us correlation functions.

\section{The Quantum Drift Model}

By using the Trotter-Suzuki expansion, the quantum dynamics is obtained by
the sequential application of unitary evolution operators that transform the
initial state in small time-steps, $\mathrm{d}t$. If the system-environment
interaction has a rate $1/\tau _{SE}$, during each interval $\mathrm{d}t$,
the particle has a probability $p=\mathrm{d}t/\tau _{SE}$ to undergo a
decoherent process and a probability $(1-p)$ to survive it \cite{GLBE1,GLBE2}%
. Let us consider a single state $\left\vert 0\right\rangle $ that may
undergo a decoherent process. Thus, after a time $\mathrm{d}t$, the coherent
amplitude is reduced by the factor $\sqrt{1-p}$ owing to decoherent
processes. In order to conserve the density, and in accordance with the
Landauer-B\"{u}ttiker picture, the wave function must include a term that
accounts for the decoherent reinjection. Thus it has a random phase $\theta $
drawn from some distribution $P_{\theta }$. We can represent both the
coherent and the incoherent contributions in the same wave function,
resembling quantum parallelism. This is,

\begin{eqnarray}
\widetilde{\psi }_{0} &=&\overset{}{\psi }_{0}^{coh.}+~\overset{}{\psi }%
_{0}^{incoh.} \\
&=&\left( \sqrt{1-p}+\lambda _{\theta }e^{\mathrm{i}\theta }\right) \overset{%
}{\psi }_{0}
\end{eqnarray}%
where $\widetilde{\psi }_{0}=\left\langle 0\right\vert \left. \widetilde{%
\psi }\right\rangle $. The cross terms cancellation in the ensemble average
is ensured by $\int P_{\theta }\sqrt{1-p}\lambda _{\theta }e^{\mathrm{i}%
\theta }d\theta =0$. In any case, the coefficient $\lambda _{\theta }$
should be chosen to account for density conservation $\left\vert \sqrt{1-p}%
+\lambda _{\theta }e^{\mathrm{i}\theta }\right\vert \equiv 1$. Thus, 
\begin{equation}
\widetilde{\psi }_{0}=e^{\mathrm{i}\beta _{0}}\overset{}{\psi }_{0},
\end{equation}%
for some random phase $\beta _{0}$.

Notice that, in a Trotter-Suzuki evolution, the phase shift $e^{i\beta _{0}}$
is actually a correction $\Delta E_{0}=\hbar \beta _{0}/\mathrm{d}t$ in the
energy of the state $\left\vert 0\right\rangle $. Equation \ref%
{Eq_Lorentzian-convolution} shows that the single level coupled to an
environment acquires an energy uncertainty $\Gamma _{\phi }$, which in a FGR
approximation, is characterized by a Lorentzian shape. This, in turn, is
assimilated to a distribution of instantaneous energy shifts $\Delta E_{0}$
drawn from the Lorentzian distribution. In our model, the correction $\Delta
E_{0}$\ is taken to be a random number that varies step-by-step to represent
the uncertainty introduced by the environment. Thus, the probability
distribution $P_{\beta _{0}}$ is 
\begin{equation}
P(\beta _{0})=\frac{1}{\pi }\frac{\Gamma _{\phi }\mathrm{d}t/\hbar }{\beta
_{0}^{2}+(\Gamma _{\phi }\mathrm{d}t/\hbar )^{2}}.
\label{eq:Prob_Lorentzian}
\end{equation}%
Therefore, the key decoherent processes are the highly improbable processes
that involve a large $\Delta E_{0}$ (the tails of the Lorentzian).

This proposal can be extended to all the levels $\ E_{n}$ of a Hamiltonian
in an arbitrary basis. In particular, in a tight-binding basis, each site
energy $E_{n}$ acquires a Lorentzian energy uncertainty $\Gamma _{\phi ,n}$
and these are perturbed with a random energy $\Delta E_{n}$. More formally,
we can define $\hat{\Sigma}$ as a diagonal operator where $\Sigma
_{n,n^{\prime }}=\Delta E_{n}\delta _{n,n^{\prime }}$. For a $N\times N$
matrix Hamiltonian $H_{0}$ we consider an effective instantaneous
Hamiltonian $\widehat{\tilde{H}}_{eff.}=\hat{H}_{0}+\hat{\Sigma}$. Thus, we
obtain the unitary evolution operator in a Trotter-Suzuki expansion,%
\begin{eqnarray}
\widehat{\tilde{U}}(\mathrm{d}t) &=&e^{-\mathrm{i}\widehat{H}_{eff}\mathrm{d}%
t/\hbar }, \\
&\simeq &e^{-\mathrm{i}\hat{\Sigma}\mathrm{d}t/\hbar }e^{-\mathrm{i}\widehat{%
H}_{0}\mathrm{d}t/\hbar }=\hat{U}_{\Sigma }(\mathrm{d}t)\hat{U}_{0}(\mathrm{d%
}t).
\end{eqnarray}%
We can define the decoherence operator as $\hat{U}_{\Sigma }=$\ $\exp [-%
\mathrm{i}\hat{\Sigma}\mathrm{d}t/\hbar ]$, that is unitary, it conserves
the density and, thus, the density does not presents jumps. However,
observables involving two-sites correlation (e.g. currents and momentum) do
have jumps. However, these last are smoothed out by taking the ensemble
average.

In summary, the prescription to include decoherence in a quantum dynamics is
to include in every time step a random correction, $\beta _{n}$, to the
phase of each local state. This correction has a distribution probability
given by the $P_{\beta _{n}}$\ of Eq. \ref{eq:Prob_Lorentzian}. Thus, the
evolution of a wave function is performed by%
\begin{equation}
\left\vert \widetilde{\psi }(t)\right\rangle =\prod_{j=1}^{N_{t}}e^{-\mathrm{%
i}\hat{\Sigma}\mathrm{d}t/\hbar }e^{-\mathrm{i}\widehat{H}_{0}\mathrm{d}%
t/\hbar }\left\vert \psi (0)\right\rangle ,
\end{equation}%
where $N_{t}=t/\mathrm{d}t$.

\section{Decoherent transport: D'Amato-Pastawski Transmittance}

Let us first test our model in the system that inspired it: decoherent
transport through a double barrier resonant tunneling device (DBRTD). In a
tight-binding scheme, this is represented by one resonant site of energy $%
E_{0}=0$ coupled to two semi-infinite leads of bandwidth $4V$ (where $V$ is 
the unit of energy), $L$ and $R$
that act as current source and drain. The tunneling amplitudes through the
barriers are $V_{L}$ and $V_{R}.$ Thus, the tight-binding Hamiltonian is

\begin{eqnarray}
\hat{H}_{0} &=&E_{0}\hat{c}_{0}^{+}\hat{c}_{0}^{{}}-V_{L}(\hat{c}_{-1}^{+}%
\hat{c}_{0}^{{}}+\hat{c}_{0}^{+}\hat{c}_{-1}^{{}})-V_{R}(\hat{c}_{1}^{+}\hat{%
c}_{0}^{{}}+\hat{c}_{0}^{+}\hat{c}_{1}^{{}}) \\
&&-\sum\limits_{n=1}^{\infty }V(\hat{c}_{n+1}^{+}\hat{c}_{n}^{{}}+\hat{c}%
_{n}^{+}\hat{c}_{n+1}^{{}})-\sum\limits_{n=-1}^{-\infty }V(\hat{c}_{n+1}^{+}%
\hat{c}_{n}^{{}}+\hat{c}_{n}^{+}\hat{c}_{n+1}^{{}}).  \notag
\end{eqnarray}

To evaluate the transmittance from a dynamical calculation, we build a
Gaussian wave packet well inside the left lead. A wide wave packet ensures a
well defined energy. The transmission coefficient is obtained by integrating
the density at the right side after the wave packet has been transmitted or
reflected. This transmittance is equivalent to the steady-state analytic
result of the Fisher and Lee formula \cite{PM01}.

\begin{figure}[ptb]
\begin{center}
\includegraphics[width=0.48\textwidth]{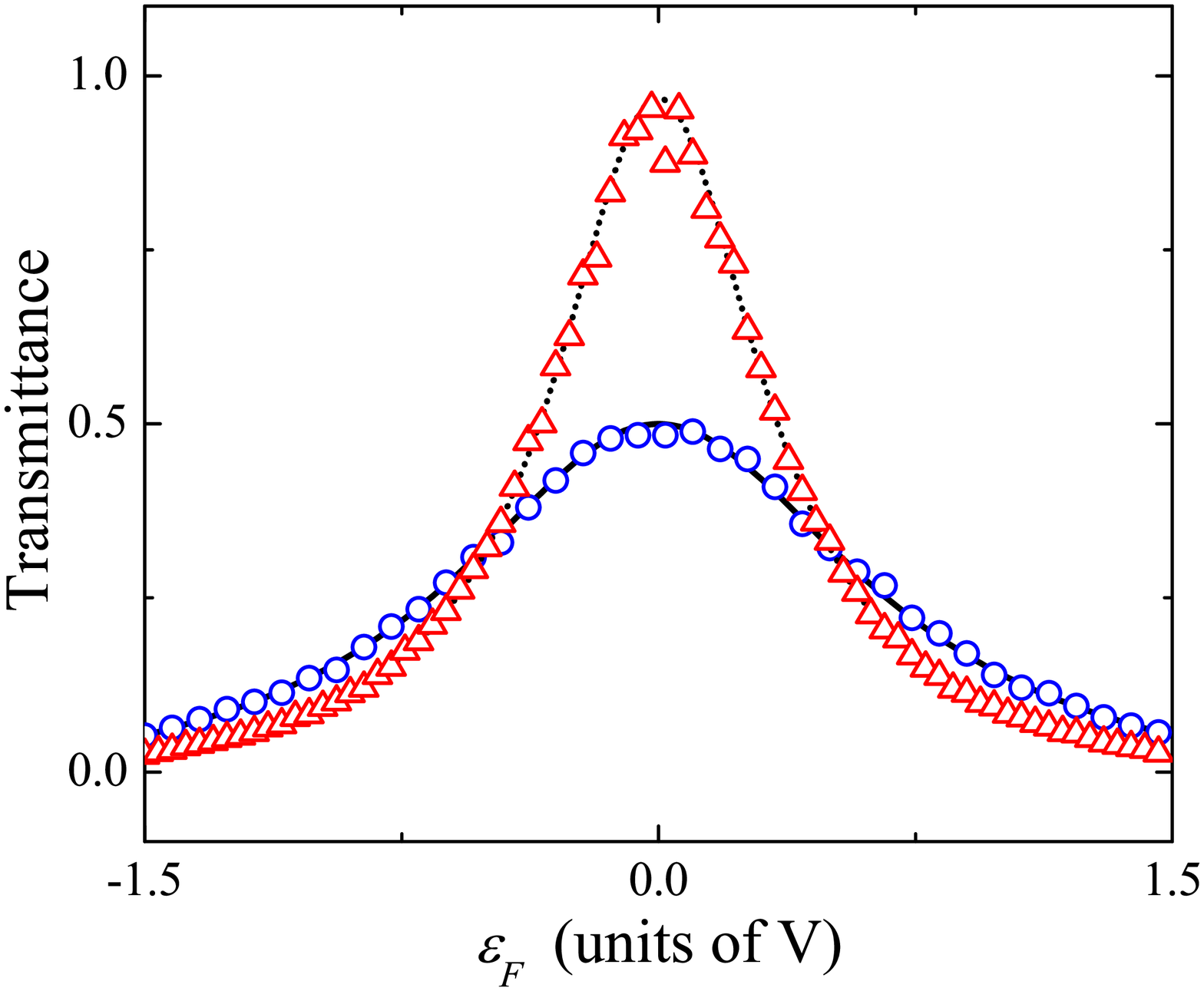}
\end{center}
\caption{(Color online) Decoherent transmittances obtained for two values of
the decoherence strength. For $\Gamma _{\protect\phi }/V=0.01$, the B\"{u}%
ttiker transmittance is plotted with the short-dot line and the
transmittance of the QD model is plotted with the red triangles for a number 
$N_{S}=5$ of realizations in the average. For $\Gamma _{\protect\phi }/V=0.3 
$, the B\"{u}ttiker transmittance is showed by the solid line and the QD
result, with the blue circles for a number of $N_{S}=50$ realizations. The
other parameters are: $V=1$,~$E_{0}=0$,~$V_{L}=V_{R}=0.15$, $E_{r}=0$}
\label{fig_transmittances}
\end{figure}

Decoherence is introduced only at the resonant level as described in the
previous section during the whole evolution. In Fig. \ref{fig_transmittances}
we compare the QD results with those resulting from the B\"{u}ttiker's
solution of Eq. \ref{eq_decoherent_transmittance}. We plot these quantities
for different decoherence strengths $\Gamma _{\phi }$. These are in a very
good agreement, made even more valuable by considering that the number
of realizations in the average was of the order of 10. This is because the
observable of transmitted density involves a spatial integration. The same
self-average is observed for the decoherent conductance in long one-dimensional (1D) wires. In
that case even a single realization is enough to reproduce the known results
in this problem \cite{DP90}.

The QD method fits the theoretical values to the desired precision. The only
difference that one might notice in certain specific cases, such as narrow peaks,
would arise from the fact that wave packets are built from states within an
energy range as a consequence of the uncertainty principle. On the other hand,
scattering theory, by using asymptotic plane waves, has no energy
uncertainty. This example is representative of a wide variety of steady
state problems that can be solved with the QD method, at only the cost of
numerical resources. Indeed, we tested QD on extended systems, a situation
where an implementation of the QJ would impose excessive local fluctuations.
We found that in QD, much as in quantum parallelism \cite{ADLP07}, the
collective observables have a tendency to self-average. This makes our model
a very promising tool to evaluate decoherent dynamics in extended systems
and in many-body problems. However, the true advantage of QD starts to be
appreciated in addressing time-dependent problems, as we do in the next
section.

\section{Quantum Dynamical Phase Transition in a Two Level System.}

Let us consider a two-level system (TLS) that describes charge or spin
dynamics, \cite{DALP07(ssc),SKVCirGi14} say states $\left\vert
0\right\rangle $ $\equiv $ $\left\vert \uparrow \downarrow \right\rangle =%
\hat{c}_{0}^{+}\left\vert vacuum\right\rangle $ and $\left\vert
1\right\rangle \equiv $ $\left\vert \downarrow \uparrow \right\rangle =\hat{c%
}_{1}^{+}\left\vert vacuum\right\rangle $, with degenerate energy $E_{0}$
and an interaction $V$ mixing them. Such simple system was seen to have a
non-trivial dynamics when one of its levels interacts with an environment of
spins: a quantum dynamical phase transition (QDPT) \cite{ADLP06}. In a QDPT
certain observables present a non-analytic dependence on the
system-environment interaction strength. The QDPT was missed in a solution
for the density matrix in the usual secular approximation of the Redfield
theory \cite{MKBE74} but showed up in a QJ variant \cite{ADLP06}. When the
TLS suffers the asymmetric interaction of an environment, the QDPT already
appears in the spectrum of the effective non-Hermitian Hamiltonian \cite%
{Rot09}. However, if the interaction is symmetrical, the QDPT only occurs in
the density matrix if positivity is ensured \cite{P2007PhysB}. Thus,
obtaining the QDPT in a model with symmetric interaction with the
environment constitutes a definitive test for the QD method.

The Hamiltonian of the TLS is 
\begin{equation}
\hat{H}_{0}=E_{0}(\hat{c}_{0}^{+}\hat{c}_{0}^{{}}+\hat{c}_{1}^{+}\hat{c}%
_{1}^{{}})-V(\hat{c}_{1}^{+}\hat{c}_{0}^{{}}+\hat{c}_{0}^{+}\hat{c}%
_{1}^{{}}).
\end{equation}%
The survival probability of an excitation with an initial state $\left\vert
\psi \left( 0\right) \right\rangle =\left\vert 0\right\rangle $, i.e. the
diagonal element of the density matrix, is 
\begin{eqnarray}
P_{00}(t) &=&\left\vert \left\langle 0\right\vert e^{-\mathrm{i}\widehat{H}%
_{0}t/\hbar }\left\vert 0\right\rangle \right\vert ^{2}, \\
&=&1/2+1/2\cos \omega _{0}t.
\end{eqnarray}%
Here, we can observe that $P_{00}(t)$ has Rabi oscillations \cite{Cohen92}
with frequency $\omega _{0}=2V/\hbar $ and period $T=\pi \hbar /V$.

Let us consider now an environment that interacts independently with each\
state with a rate described by the Fermi golden rule $1/\tau _{SE}=2\Gamma
_{\phi }/\hbar $, where, $\Gamma _{\phi }$ is the energy uncertainty of the
level. Physical implications of this model are discussed in the next section
in the context of a spin system. The numerical evolution of the TLS is now
performed by choosing $\Gamma _{\phi }$\ as the width of the Lorentzian
distribution.

Here we will compare our QD method, where the decoherent survival
probability is

\begin{equation}
\tilde{P}_{00}(t)=\left\vert \left\langle 0\right\vert
\prod_{n=1}^{N_{t}}e^{-\mathrm{i}\hat{\Sigma}_{n}\mathrm{d}t/\hbar }e^{-%
\mathrm{i}\widehat{H}_{0}\mathrm{d}t/\hbar }\left\vert 0\right\rangle
\right\vert ^{2},
\end{equation}%
with the analytic solution of the GLBE. This last was analytically solved
for this problem in Refs. \cite{GLBE1,ADLP06,P2007PhysB}, giving for the
survival probability%
\begin{equation}
\tilde{P}_{00}(t)=\frac{1}{2}+\frac{1}{2}e^{-\Gamma _{\phi }t/\hbar }\left[
\cos \left( \omega t\right) +\frac{\Gamma _{\phi }}{2\omega }\sin \left(
\omega t\right) \right] .  \label{eq_prob_underdamped}
\end{equation}%
Thus, the oscillations of both the diagonal and non-diagonal elements of the
density matrix oscillate with a frequency $\omega $, which is lower than the
Rabi frequency $\omega _{0}$ according to 
\begin{equation}
\omega =\sqrt{\omega _{o}^{2}-\left( \Gamma _{\phi }/\hbar \right) ^{2}}.
\label{eq.frec_real}
\end{equation}%
This evidences that the oscillation frequency of a TLS exhibits a
non-analytic behavior. The frequency $\omega $ takes real values provided
that $\Gamma _{\phi }/\hbar <\omega _{o}$ (underdamped regime). Beyond this
value, i.e., for $\Gamma _{\phi }/\hbar >\omega _{o}$ (overdamped regime), $%
\mathit{Re}(\omega )\equiv 0$ and thus the oscillations are fully
suppressed, and $\tilde{P}_{00}(t)$ is the sum of two exponential decays:

\begin{equation}
\tilde{P}_{00}(t)=\frac{1}{2}-\frac{\gamma _{2}}{2(\gamma _{1}-\gamma _{2})}%
e^{-\gamma _{1}t}+\frac{\gamma _{1}}{2(\gamma _{1}-\gamma _{2})}e^{-\gamma
_{2}t},  \label{eq_prob_overdamped}
\end{equation}%
where the decay rate $\gamma _{1(2)}$ is 
\begin{equation}
\gamma _{1(2)}=\frac{1}{\hbar }\left( \Gamma _{\phi }\pm \sqrt{\Gamma _{\phi}^{2}-(\hbar \omega _{o})^{2}}\right) .  \label{eq_rates_overdamped}
\end{equation}%
Note that, at short times, $\tilde{P}_{00}(t)$ is always of the form $%
1-\omega _{o}^{2}t^{2}/4=1-V^{2}t^{2}/\hbar^{2}$, which is characteristic of a quantum
evolution without perturbations. This is because, at short times, the
environment interplay has a small cumulative effect on the survival
probability, which is still determined by the unperturbed quantum dynamics.
In a strongly decoherent regime, $\Gamma _{\phi }/\hbar \gg \omega _{o}$,
the decay rates tend to $\gamma _{1}\simeq 2\Gamma _{\phi }/\hbar $, $\gamma
_{2}\simeq \hbar \omega _{o}^{2}/2\Gamma _{\phi }$, defining a short-time
decay rate, $\gamma _{1}$, and a rate $\gamma _{2}$, that dominates the long
times as $\tilde{P}_{00}(t)\propto e^{-\gamma _{2}t}$. Both exponential
terms are needed to obtain the whole evolution. An equivalent solution for
the QDPT may be obtained by considering models for environmental noise \cite%
{Altshuler14}.

\begin{figure}[tbh]
\begin{center}
\includegraphics[width=0.48\textwidth]{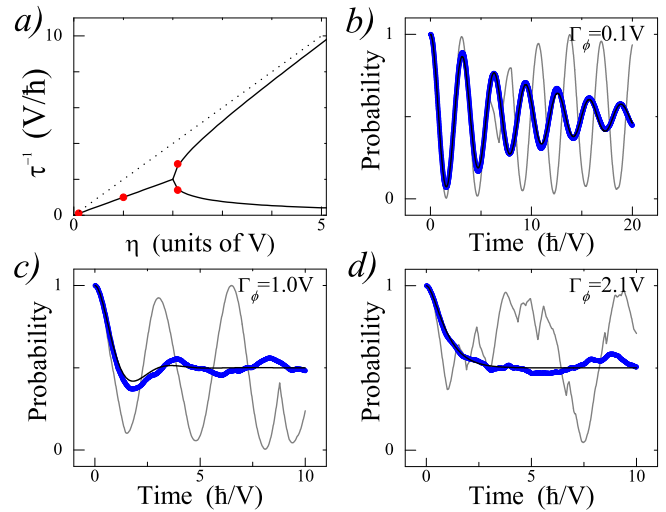}
\end{center}
\caption{(Color online) $(a)$ Theoretical rates of decay predicted by the
Eqs. \protect\ref{eq_prob_underdamped} and \protect\ref{eq_rates_overdamped}%
. Red points indicate the rates of the GLBE solution for $(b)-(d)$.
The QDPT occurs as a bifurcation in the rates for the critical value $\Gamma
_{\protect\phi }/V=2$. The dotted line represents the asymptotic value $%
2\Gamma _{\protect\phi }/\hbar $. In $(b)-(d)$, we show the survival
probabilities $\tilde{P}_{00}(t)$ (thick blue line), the GLBE solutions
(thin black line), and individual realizations (thin gray line) in the
underdamped regime, in $(b)$ with $\Gamma _{\protect\phi }/V=0.1$ and $(c)$
with $\Gamma _{\protect\phi }/V=1$, and in the overdamped regime, $(d)$ with $%
\Gamma _{\protect\phi }/V=2.1$. Individual realizations tend to preserve the
oscillations. These do not have jumps in the densities but in the slopes.
The ensemble average is taken over $N_{S}=100$ realizations. For $\tilde{P}%
_{00}(t)\simeq 1/2$ are clear the typical fluctuations of the order of $1/%
\protect\sqrt{N_{S}}$. The average $\tilde{P}_{00}(t)$ tends to the GLBE
solution by increasing $N_{S}$.}
\label{fig_rates_dynamics}
\end{figure}

In Fig. \ref{fig_rates_dynamics}$(a)$ the decay rates are shown. The QDPT is
manifested as the bifurcation in these rates. The fitted rates are shown as
points superposed to the theoretical curve. In Fig. \ref{fig_rates_dynamics}$%
(b)$, we show the Rabi oscillations of the average survival probability, $%
\tilde{P}_{00}(t)$, which are exponentially attenuated with $\tau
^{-1}\simeq \Gamma _{\phi }/\hbar =1/2(\tau _{SE})^{-1}$. This is the most
common example of decoherence in TLS's. The $\tilde{P}_{00}(t)$ and the
fitted decay rates match perfectly the GLBE solution. A single realization
of QD method is also shown. Notice that there are no \textquotedblleft
jumps\textquotedblright\ in the survival probability, and an
oscillatory-like behavior dominates the whole evolution. In Fig. \ref%
{fig_rates_dynamics}$(c)$ we show the $\tilde{P}_{00}(t)$ in the underdamped
regime for a value of $\Gamma _{\phi }=1V$, where, the oscillatory behavior
is small. In Fig. \ref{fig_rates_dynamics}$(d)$ we show $\tilde{P}_{00}(t)$
in the overdamped regime, for $\Gamma _{\phi }=2.1V$. We can identify the
initial quadratic behavior and, at large times, the exponential decay with
the rate $\gamma _{2}\propto 1/\Gamma _{\phi }$. The larger $\Gamma _{\phi }$%
, the slower $\tilde{P}_{00}(t)$ decays. This is a signature of the \textit{%
quantum Zeno effect} in which the system is continuously perturbed freezing
its evolution close to the initial condition. By increasing $\Gamma _{\phi }$%
, each single realization seems to be a stochastic process while preserving
the quadratic initial starting. Single realizations do not present jumps in
the density but in the correlations, seen as sudden changes on the slopes.

The actual dynamics of the observables emerge after ensemble averaging. As
long as the survival probability is not too close to $1/2$,\ a fair
representation of $\tilde{P}_{00}(t)$ is obtained with about $N=100$\
realizations as shown in Fig. \ref{fig_rates_dynamics}. Strongly damped
cases evidence the typical fluctuations of random numbers where the
observables have a precision of $1/\sqrt{N}$. In these cases, individual
systems maintain a substantial oscillation whose slopes can be strongly
discontinuous. Thus to obtain coincidence with the exact theoretical values
within the graphical resolution (say $1\%$) one needs about $N=10 \ 000$
realizations. We tested a binary and Gaussian phase drift distributions and
the fluctuations and their influence on the precision of the ensemble
averages persist. However, this is hardly a limitation if one is still far
from the asymptotic values or when one addresses global observables.

\section{Quantum Jumps vs. Quantum Drift in a many-spin dynamics}

Here, we will further assess the differences between the QJ and QD in a
situation of actual experimental relevance where the density matrix approach
is clearly restrictive: that of many-spin dynamics. We will address the
decoherent dynamics of this problem, which is non-trivial in terms of the 
density matrix. We are interested in the
dynamics of a local spin excitation in a system of $M$ interacting spins $%
1/2$. Let us say that the state at $t=0$ is given by the density matrix: 
\begin{equation}
\hat{\rho}_{0}=\frac{1}{2^{M}}(\mathbf{\hat{I}}+2\hat{S}_{1}^{z}),
\label{inistate}
\end{equation}%
which describes that spin $1$ is up-polarized. At very high temperature the
other spins are not polarized at all, i.e. $tr[\hat{S}_{1}^{z}\hat{\rho}%
_{0}]=\frac{1}{2}$ and $tr[\hat{S}_{i}^{z}\hat{\rho}_{0}]=0$ $\forall i\neq
1 $. In order to be more specific, let us consider the particular case of a
linear chain with $M=5$ that was addressed theoretically and experimentally
in NMR and where a decoherent calculation is lacking \cite%
{Ernst97,Past96CPL,P95PRL}. There, the effective Hamiltonian is reduced to
nearest-neighbors planar (or $XY$) interactions. The Hamiltonian, using the
spin lowering $\hat{S}_{i}^{-}$ and raising $\hat{S}_{i}^{+}$ operators is: 
\begin{eqnarray}
\hat{H}_{chain} &=&\sum_{i=1}^{M-1}J_{i,i+1}(\hat{S}_{i}^{+}\hat{S}%
_{i+1}^{-}+\hat{S}_{i+1}^{+}\hat{S}_{i}^{-})  \notag \\
&&+\sum_{i=1}^{M}\hbar \Omega _{i}(\hat{S}_{i}^{+}\hat{S}_{i}^{-}-\frac{1}{2}%
).
\end{eqnarray}%
The first term is the $XY$ Hamiltonian, accounting for the couplings to
nearest-neighbours. We will take $J_{i,i+1}=J$ as the unit of energy.
The second one is the Zeeman Hamiltonian, where the
precession frequencies are $\Omega _{i}=\omega _{0}$, and $\omega _{0}$\
is the Larmor frequency in the external magnetic field. As predicted by the
coherent calculation, the local excitation $\hat{\rho}_{0}$ was seen to
propagate as a spin wave through the molecular chain and returns to the
initial site in the form of a mesoscopic echo (ME). It is precisely the wave
packet behavior that makes these systems promising as quantum channels \cite%
{CRC07,Bose03,CrisDEL07,Zwick11}. However, the experiments show that
these spin waves decohere and attenuate as time pass by. Thus, here we
consider that each spin is perturbed by a local environment that acts as a
fluctuating Zeeman field. Thus $\Omega _{i}=\omega_{0}+\delta \omega _{i}$,
where $\delta \omega _{i}$ fluctuates with time.

\begin{figure}[ptb]
\begin{center}
\includegraphics[width=0.38\textwidth]{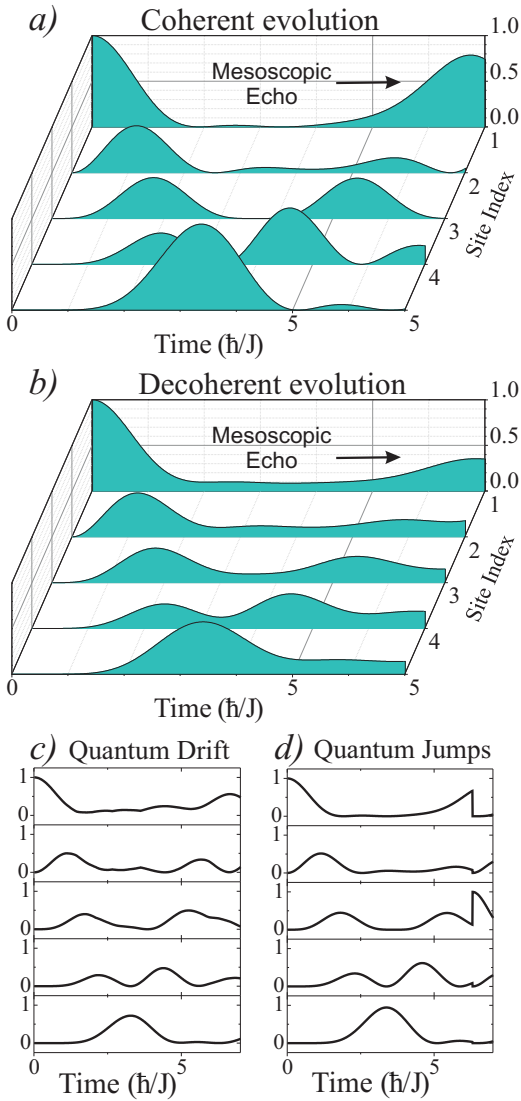}
\end{center}
\caption{(Color online) Density as function of time at each spin for $(a)$
the coherent evolution, $(b)$ an average evolution with a finite coherence time $%
T_{coh}=3\hbar/J$, $(c)$ a single realization of the QD model, and $(d)$ a single
realization of the QJ model. When the spin wave reaches the edge of the chain, it is reflected
constituting the mesoscopic echo. Note that the curves for QD are
smooth in density whereas those for QJ present a sudden change when the jump
is produced. }
\label{fig_ernst}
\end{figure}

We will solve this problem by resorting the Jordan-Wigner transformation 
\cite{DPL02PhysB} which in this case looks $\hat{S}_{i}^{+}%
\longleftrightarrow \hat{c}_{i}^{+}$, $\hat{S}_{i}^{-}\longleftrightarrow 
\hat{c}_{i}^{{}}$ and $\hat{S}_{i}^{z}\longleftrightarrow \hat{c}_{i}^{+}%
\hat{c}_{i}^{{}}-1/2$ . Thus, this many-spin system can be reduced to a
one-body problem. This transformation remains valid when considering the
random fluctuations in the Zeeman field. Then, it is clear that these
produce local energy fluctuations $\hbar \delta \omega _{i}$ that show up as
decoherence in the spin dynamics. Note that the fluctuations are naturally
described within the QD prescription and, conversely, the random phases of
QD have a direct physical meaning in this problem.

In Figs. \ref{fig_ernst}$(a)$ and \ref{fig_ernst}$(b)$ we compare the coherent and
decoherent evolutions of the spin wave. The local excitation travels from
site $1$ to the edge of the chain and there it is reflected and returns to
the initial spin as a ME. We use a decoherence time of about $\ 3\hbar /J$,
consistent with the experimental observation.\cite{Ernst97} In Figs. \ref%
{fig_ernst}$(c)$ and \ref{fig_ernst}$(d)$ we show the local density for a single realization
for both the QD and QJ methods. The QD has a smoother profile whereas the QJ
resembles the coherent evolution until the jump is produced.

In similar problems, the master equation for the density matrix could be
used to obtain the decoherent dynamics \cite{ZnidHv13}. However, with regard to numerical calculations, it is very demanding. Thus, the QJ
method, which is based on a stochastic wave function, has proved to be more
convenient in addressing large systems. \cite{Petr97} Indeed, it is faster
and less demanding to perform an average over many realizations of the QJ
than computing the whole density matrix. Since our QD method is also based
in a stochastic wave function, for large enough systems the QD must become more
convenient than a density matrix approach. Thus, we shall compare the QD
with the QJ.

\begin{figure}[ptb]
\begin{center}
\includegraphics[width=0.5\textwidth]{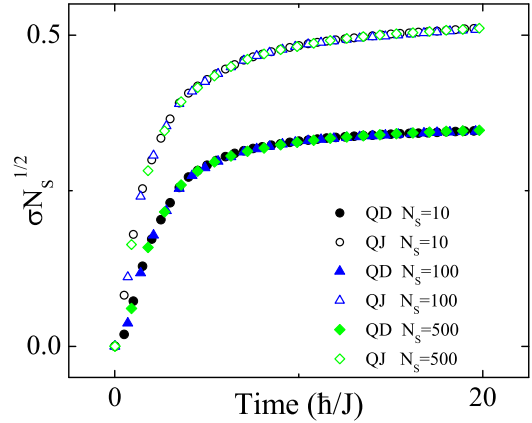}
\end{center}
\caption{(Color online) The standard error $\protect\sigma $ times $\protect%
\sqrt{N}_{s}$ as function of the evolution time, for $N_{s}=10,100$, and $500$%
. Note that $\protect\sigma N_{s}^{1/2}$ does not depends on $N_{s}$ but in
the QD or QJ method. Note that, for the same $N_{s}$, $\protect\sigma ^{QJ}/%
\protect\sigma ^{QD}\approx 2$.}
\label{fig_RMS}
\end{figure}

We will look at the convergence to the average decoherent dynamics 
for both QJ and QD methods. Due to the discontinuous nature of QJ, 
we expect that the difference with the mean dynamics is greater for QJ than for
QD (see Figs. \ref{fig_ernst}$(c)$ and \ref{fig_ernst}$(d)$). This can be quantified in terms of the time average of the standard
error,%
\begin{equation}
\sigma ^{(QD,QJ)}=\sqrt{\frac{1}{N_{s}~T}\int_{0}^{T}\mathrm{d}%
t\sum_{i=1}^{N_{s}}\left( \rho _{i,i}^{(QD,QJ)}(t)-\bar{\rho}%
_{i,i}(t)\right) ^{2}}.
\end{equation}%
Here, $T$ is the evolution time and $\bar{\rho}_{ii}(t)=\overline{\left\vert
i\right\rangle \left\langle i\right\vert }$ is the mean density at time $t$
at site $i$ obtained from an average over $N_{s}=10^{5}$ realizations, where both
QJ and QD converge to the the same average dynamics with 
negligible error. In Fig. \ref{fig_RMS} we show $\sigma N_{s}^{1/2}$ for
different values of $N_{s}$ as function of the evolution time $T$. Thus,
this result evidences the general tendency of the QD method to present a
smaller deviation from the mean than the QJ. Indeed, for each $N_{s}$, $%
\sigma ^{QJ}$ is greater than $\sigma ^{QD}$ by a factor of almost $2$.
Thus, to converge with a given $\sigma $, QD needs to perform roughly the
half of realizations than QJ.

\section{Loschmidt Echo}

Following the logic of the two previous sections, one would be tempted to
assign the meaning of decoherence to the decay of the oscillations. However,
it has been clarified that a way to filter away the intrinsic dynamics from
the observable is by using the Loschmidt echo (LE) \cite{JalP01}. This is
the amount of excitation recovered after a time-reversal procedure
implemented in presence of an environment. The advantage is that the LE
codifies in a local observable the losses of non-local correlations. As in
NMR experiments, this consists in changing the sign of the acting
Hamiltonian $\widehat{H}_{0}\longrightarrow -\widehat{H}_{0}$. By using the
Trotter-Suzuki expansion, the LE can be defined as:

\begin{eqnarray}
M_{00}(2t) &=&\left\vert \left\langle 0\right\vert
\prod_{m=N_{t}+1}^{2N_{t}}e^{+\mathrm{i}\widehat{H}_{0}\mathrm{d}t/\hbar
}e^{-\mathrm{i}\hat{\Sigma}_{m}\mathrm{d}t/\hbar }\right. \times  \notag \\
&&\left. \prod_{n=1}^{N_{t}}e^{-\mathrm{i}\hat{\Sigma}_{n}\mathrm{d}t/\hbar
}e^{-\mathrm{i}\widehat{H}_{0}\mathrm{d}t/\hbar }\left\vert 0\right\rangle
\right\vert ^{2}
\end{eqnarray}%
where $N_{t}=t/\mathrm{d}t$, and $\hat{\Sigma}_{n}$ is the perturbation's
Hamiltonian. Note that, whereas the sign of the Hamiltonian changes,
the perturbation remains with the same sign.

\begin{figure}[ptb]
\begin{center}
\includegraphics[width=0.48\textwidth]{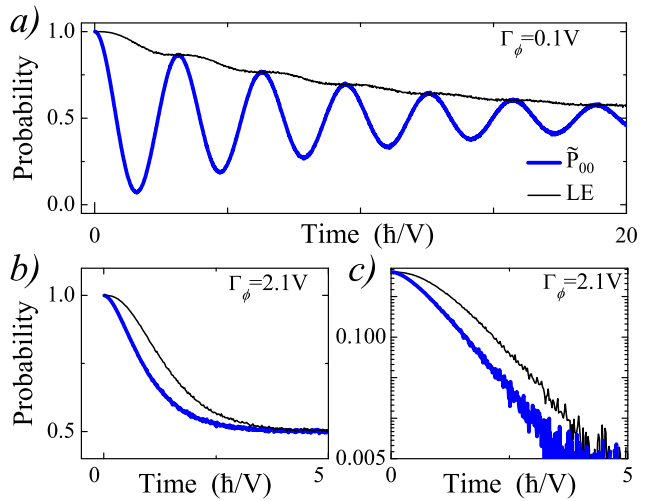}
\end{center}
\caption{(Color online) Survival probability (thick blue line) and the
Loschmidt echo (thin black line) as function of the time of action of the
environment for: $(a)$ $\protect\eta /V=0.1$ and $(b)$ $\protect\eta /V=2.1$,
averaged over $N_{S}=10000$ realizations. $(c)$: the same as in $(b)$ but
in log scale. The rate of decay of the LE, $1/\protect\tau _{SE}=1.12V/\hbar 
$, is smaller than the minimum decay rate of the $\tilde{P}_{00}$, $1/%
\protect\tau _{2}=1.46V/\hbar $.}
\label{fig_Loschmidt}
\end{figure}

In Fig. \ref{fig_Loschmidt}$(a)$, we show the survival probability $%
\widetilde{P}_{00}(t)$ and the Loschmidt echo $M_{00}(t)$ in the underdamped
regime as function of the time of interaction with the environment, $t$.
Surprisingly, in the underdamped regime, $M_{00}$ is not a simple
exponential but has plateaus whenever the reversal starts while the system
is at $\left\vert 0\right\rangle $ or $\left\vert 1\right\rangle $. On the
contrary, $M_{00}$ suffers a maximal decay if the reversal starts when the
system is at a superposition state $\left[ \left\vert 0\right\rangle \pm
\left\vert 1\right\rangle \right] /\sqrt{2}$. When the density is placed on
one site, decoherent interactions act as a change in the global phase, which
does not destroy the phase correlations between $\left\vert 0\right\rangle $
and $\left\vert 1\right\rangle $. Notice that the homogeneous exponential
decay of the survival probability does not discriminate on the initial
state, while the LE does. In any case, if one should define a overall rate $%
\gamma _{\phi LE}\equiv 1/\tau _{\phi LE}$ from the LE, it would coincide
with that observed in the oscillation decay, i.e., $\gamma _{\phi LE}\simeq
\Gamma _{\phi }/\hbar =1/2(\tau _{SE})^{-1}$. Thus, the LE gives a rationale
to the $1/2$ factor: decoherent processes are effective on one half of the
dynamical cycle.

The overdamped regime is shown in Fig. \ref{fig_Loschmidt}$(b)$ and in a log
scale in Fig. \ref{fig_Loschmidt}$(c)$ . This last clarifies the different decay rates and the
difficulties to obtain probabilities around 1/2 from a limited number of
realizations. The LE has a wider plateau than the survival probability at
short times. For the same period of action of the decoherent processes, the
LE signal is higher than the survival probability. This fact is consistent
with that, in order for decoherent processes to be effective, the dynamics
must first build up the superposition state. The decay rate of the LE does
not fit with Eq. \ref{eq_rates_overdamped}. By using a single exponential,
the decay rate is $1/\tau _{\phi LE}=1.12V/\hbar $, which is slightly
smaller than the minimum rate of $\tilde{P}_{00}$, $\gamma _{2}=\Gamma
_{\phi }-\sqrt{\Gamma _{\phi }^{2}-(\hbar \omega _{o})^{2}}=1.46V$. This indicates
that the LE gives more weight to the less correlated short-time processes
where the strong interaction with the environment does not allow us to create
the correlations and thus, it should have a slower decay.

\section{Final Discussion}

We developed and implemented a stochastic model to include decoherent
processes in quantum dynamics. Inspired in the B\"{u}%
ttiker-D'Amato-Pastawski description of decoherent transport, it boils down
to a wave function that undergoes smooth stochastic drifts in a local
basis.

Unlike the quantum jumps (QJ) approach, no collapses of the wave function
occur and phase shifts are introduced in a unitary dynamics. Thus, in QD
jumps can only appear in the correlations functions, not in the local
densities. Being an appealing conceptual framework, with a clear physical
meaning, our QD model results particularly adapted to deal with extended
system. Besides, it admits further extensions, ranging from the evaluation
of currents in transport set-ups to the representation of specific many-body
interactions.

Using numerical calculations, we proved that our dynamical model is in a
full agreement with the decoherent-steady-state conductances through the
resonant state $\left\vert 0\right\rangle $ of a decoherent quantum dot,
even resorting to a quite restricted ensemble average. For steady-state
transport in extended systems, a QD evaluation of the wave function is,
by construction, more efficient than density matrix approaches \cite%
{ZnidHv13}.

In spin systems, the physical foundation of the QD model becomes evident. 
Decoherence associated to the fluctuation of the local energy is a natural
ingredient associated with the fast fluctuation of the local Zeeman fields.
Thus, we tested the dynamical properties of QD model by applying it to a two-spin system and a five spin system in presence of decoherence. The first is
a two-level system (TLS) that oscillates among $\left\vert 0\right\rangle $ $%
\equiv $ $\left\vert \uparrow \downarrow \right\rangle $ and $\left\vert
1\right\rangle \equiv $ $\left\vert \downarrow \uparrow \right\rangle $.
There, a non-trivial quantum dynamical phase transition shows up, which was observed in NMR and
obtained from the solution of the generalized Landauer-B\"{u}%
ttiker equations \cite{ADLP07,DALP07(ssc)}. We recovered not only the
exponential damping of the oscillations at low rates of interaction with the
environment but also the bifurcation of the decoherence rates at a critical
interaction strength. The evaluation of the decoherent dynamics of a
five-spin system is also done in connection with NMR experiments \cite%
{Ernst97}. By using rates consistent with the experiment, we show the
robustness of the mesoscopic echo under decoherence. We also show that, for
a given tolerance error for an observable, the QD method demands about half the
realizations than QJ.

By evaluating decoherence in the TLS through Loschmidt echo (LE), we found
that the pure states $\left\vert 0\right\rangle $ $\equiv $ $\left\vert
\uparrow \downarrow \right\rangle $ and $\left\vert 1\right\rangle \equiv $ $%
\left\vert \downarrow \uparrow \right\rangle $ are quite robust against
local perturbations of the environment. In contrast, the LE, and hence
coherence, decays faster when the system is in a superposition state $\left(
\left\vert \uparrow \downarrow \right\rangle \pm \left\vert \downarrow
\uparrow \right\rangle \right) /\sqrt{2}$. These results are in agreement
with the general trend recently observed in spin systems through NMR \cite%
{Sanchez14}.

In summary, we proposed a QD model that provides a stochastic unitary
dynamics of the wave function. Observable evaluation of observables through
QD and QJ are naturally parallelizable and thus they result in being more scalable
than density matrix methods\cite{Petr97}. This quality, which adds to the
intrinsic physical significance, should make the QD method a suitable tool
to address dynamical observables in both extended one-body and many-body
systems.

\acknowledgments The authors acknowledge F. Pastawski and R. Bustos-Marun
for valuable discussions. This work was performed with the financial support
of CONICET, ANPCyT, SeCyT-UNC and MinCyT-Cor.

\bibliographystyle{ieeetr}
\bibliography{./bibliografia_WOurl}

\begin{thebibliography}{10}
\bibitem{attenuation}
An explicit evaluation of the energy dependence of $\Lambda (\varepsilon
  )$\textit shows that attenuation is not much effective for the small
  off-resonance tunneling, as the electrons\ with energies away from the
  resonance spent a very short time at this interacting state.
\end{thebibliography}

\end{document}